\documentstyle[preprint,aps,version2]{revtex}
\begin{document}
\draft
\begin{title}
Small-amplitude excitations in a  deformable discrete \\
nonlinear Schr\"{o}dinger equation.
\end{title}
\author{V.V. Konotop\thanks{Permanent address:
Department of Physics and Center
of Mathematical Sciences, University of
Madeira, Pra\c{c}a do Munic\'{\i}pio,
Funchal, P-9000,  Portugal. Electronic address: konotop@dragoeiro.uma.pt}
and M. Salerno\thanks{also Istituto Nazionale di Fisica della Materia
(INFM) unita' di Salerno. Electronic address: salerno@vaxsa.csied.unisa.it}}
\begin{instit}
Department of Theoretical Physics, University of Salerno,
I-84100, Salerno, Italy

\end{instit}

\begin{abstract}

A detailed analysis of the small-amplitude solutions of a
deformed discrete nonlinear Schr\"{o}dinger equation is performed.
For generic deformations the system  possesses
"singular" points which split the infinite chain in a number
of independent segments. We show that small-amplitude dark solitons
in the vicinity of the singular points are described by the
Toda-lattice equation while away from the singular points are
described by the Korteweg-de Vries equation.
Depending on the value of the deformation parameter and of the
background level several kinds of solutions are possible.
In particular we delimit the  regions in the parameter space in which
dark solitons are stable in contrast with regions in which
bright pulses on  nonzero background are possible. On the
boundaries of these regions we find that  shock waves and rapidly
spreading solutions may exist.
\end{abstract}

\section{introduction}

The general discrete nonlinear Schr\"{o}dinger  equation (GDNLS)
\begin{equation}
\label{gdnls}
i\dot{q}_n +(1 + \eta |q_n|^2)
(q_{n-1}+q_{n+1}-2q_n)+ 2 (\omega_n + |q_n|^2)q_n=0.
\end{equation}
was introduced in ref.\cite{S} as a generalization
of the simple tight-binding linear  Schr\"{o}dinger model for
the dynamics of quasiparticles in a molecular crystal. In this equation
$q_n$ represents the complex mode amplitude of the molecular
vibration at site $n$, $\omega_n$  
is the onsite frequency of
the vibration while the nonlinear terms arise, in the adiabatic and small
field approximation,  as the result of the interaction of the
quasiparticle with the lattice.
From a mathematical point of view this system represents a norm preserving
deformation of the diagonal discretization of the
nonlinear Schr\"{o}dinger equation (DNLS).
The presence of the deformation parameter $\eta\in\Re$ in the system
allows one to study, both at the classical
and at the quantum level, the interplay between onsite-intersite
interactions as well as integrability-nonintegrability and
discrete-continuum properties \cite{MS,ESSE,CBG,KS,kon3,CBGS,HRTG}.
For $\eta=1$ (off-diagonal nonlinearity) the GDNLS
reduces to the integrable Ablowitz-Ladik (AL) model  with exact soliton
solutions while for $\eta=0$  (onsite nonlinearity)
it gives  the nonintegrable DNLS system.
For intermediate values of $\eta$ it allows all possible
splitting of the nonlinearity among three adjacent sites (equal
splitting is at $\eta=2/3$).
The properties of the localized mode solutions and of the plane
waves (modulational instability) of eq.(1) were studied in a
series of publications \cite{CBG,KS}. Bloch oscillations of
bright solitons in the external electric field ($\omega_n=\gamma n$, $\gamma$
being a constant)  at $\eta=1$ were also considered
in \cite{kon3} and in the case $0<\eta\leq 1$ in \cite{CBGS} while 
the spatial properties and the existence of breather-like impurity
modes were reported in refs.\cite{HSGT}, \cite{HRTG}. 
All these studies dealt mainly with
the attractive case (i.e. $\eta>0$) at $\omega_n=0$, in which brigth solitary
pulses are stable.

The aim of the present paper is to investigate the properties
of the small amplitude solution of the GDNLS equation in the repulsive case
$\eta<0$. In the following we refer  to this as the stable case
since the plane wave solutions  corresponding to the center of
the Brillouin zone (BZ) are modulationally stable.
To this end we consider   $\eta=-\epsilon$ with $0 <\epsilon<1$ and  
fix ${\omega}_n={\rho}^2$ with $\rho$ a constant 
which will be associated
with the amplitude of $q_n$ either at the infinity
or at the singular points (see below). 

The GDNLS is then written as
\begin{equation}
\label{e1}
i\dot{q}_n +(1-\epsilon |q_n|^2)
(q_{n-1}+q_{n+1}-2q_n)+2(\rho^2-|q_n|^2)q_n=0.
\end{equation}
The inverse scattering technique for (\ref{e1}) at
$\epsilon=1$ (stable AL)  has been developed in \cite{VK}, its dark
soliton solution  has been found in \cite{KV} and Bloch oscillations of the
dark soliton in a constant electric field were reported in \cite{kon6}
(note that we have changed the sign in 
front to the onsite nonlinearity in eq.\ref{e1} just to 
have $\epsilon=1$ as the AL limit).  

Though many aspects of the behavior of  solutions of
(\ref{e1}) at $\epsilon=1$ are very similar to the behavior
of their counterparts in the continuum limit
(see e.g. \cite{FT}) it has been shown in
\cite{HS,CKVV} that there are drastic differences between the discrete
and continuum dynamics when the amplitude  of a site excitation
approaches one. In this case
 the dispersive term becomes zero and the given site
decouples from its  neighbors. It is a direct consequence of the explicit form
of (\ref{e1}) that at $\epsilon <1$ a site must have an amplitude
$\rho=\epsilon^{-1/2}$ to become decoupled.
 When this happens we call the corresponding  point
 a singular point.  Thus
singular points, splitting
the infinite chain in a number of independent segments, exist
for the GDNLS for arbitrary nonzero values of the deformation parameter.
In the present paper, by performing a multiple scale expansion of the
GDNLS around singular points,
we show the existence of dark solitons which (in the small-amplitude
approximation) are described by the Toda-lattice equation. Away
from singular points we find that
dark solitons can exist in certain regions of parameter space $(\epsilon,\rho)$
and in the small-amplitude limit are described by
solitons of the Korteweg-de Vries (KdV) equation. In addition we find a region in the parameter
space in which dark solitons are unstable and bright pulses on
nonzero background are stable. On one of the boundaries of this region
we find, quite surprisingly, that the system becomes effectively
dispersionless and
the formation  of shock waves becomes possible. The second boundary corresponds
to effectively linear regime where  initially localized pulse spreads out
quite rapidly.
The analytical results, derived in terms of a
multiple scale expansion, are found in good
agreement  with direct numerical simulations of the GDNLS even behind
the limit of validity of the small amplitude approximation.

The paper is organized as follows. In  section II we
study singular point solutions of the GDNLS by considering  the case
of a single site between two singular points.
In section III we examine
small-amplitude dark solitons in the
neighborhood of the singular points in terms of the Toda chain.
In  section IV we consider a multiple scale expansion of
the dynamics of chain excitations when the amplitude
of the background is less than the amplitude of the
singular points. Finally we briefly discuss the regime of zero effective
nonlinearity and the
possibility of shock wave solutions in the GDNLS.
In the conclusions we summarize the main results of the paper.

\section{Singular points and one-site dynamics}

Let us consider equation  (\ref{e1}) subject to the  non-zero
boundary conditions
\begin{equation}
\label{e3a}
q_n\rightarrow\rho e^{\pm i\vartheta}e^{-ikn+\omega (k)t},\,\,\,\,\mbox{at
$n\rightarrow\pm\infty$}
\end{equation}
where  $k$ denotes the  wave vector from the first Brillouin zone
(BZ): $k\in[-\pi,\pi]$,
$\omega (k)=4(1-\epsilon\rho^2)\sin^2(k/2)$
is the  frequency of the background and $\vartheta$ is a constant phase. It
 possess the following integral of motion
\begin{equation}
\label{e4}
I=\sum_{n=-\infty}^{\infty}\left\{\epsilon(
q_n\bar{q}_{n-1}+
q_{n-1}\bar{q}_{n})+2(1-\epsilon)(|q_n|^2-\rho^2)\right\}.
\end{equation}
 As  mentioned above, if at some site $n$ the amplitude of oscillations
is $\epsilon^{-1/2}$, and hence
\begin{equation}
\label{e5}
q_n=\epsilon^{-1/2}e^{i\omega_0 t-ikn+i\vartheta}
\end{equation}
with
\begin{equation}
\label{add3}
\omega_0=2\left(\rho^2-\epsilon^{-1}\right),
\end{equation}
the evolution of  $q_{n-1}$ becomes
disconnected from $q_{n+1}$.

So, for two  singular points
placed on sites $l_{-}$ and $l_+$ (being $l_{\pm}$ integers),  there
is an integral of motion
\begin{equation}
\label{e5a}
I_{l_-,l_+}=\sum_{n=l_-+1}^{l_+}\left\{\epsilon(
q_n\bar{q}_{n-1}+
q_{n-1}\bar{q}_{n})+2(1-\epsilon)|q_n|^2\right\}
\end{equation}
associated with the dynamics of the segment between $l_-$ and $l_+$, which
naturally follows from (\ref{e4}).

The integral $I_{l_-,l_+}$ allows us to find an explicit solution in the case
in which  one point is placed between two singular points.
For the sake of definiteness let us assume that $l_{\pm}=\pm 1$,
with
\begin{equation}
\label{e5b}
q_{\pm 1}=\epsilon^{-1/2}\exp(i\omega_0 t\pm i\vartheta)
\end{equation}
being singular points.
Representing the solution of the middle point as
\begin{equation}
\label{e6a}
q_0=\epsilon^{-1/2}\nu \exp(i\omega_0 t)
\end{equation}
one arrives to the following expression for the integral of motion
\begin{equation}
I_{-1,1}=2\zeta\cos\vartheta+\frac{1-\epsilon}{\epsilon}(1+\zeta^2+\xi^2).
\end{equation}
Here $\zeta$ and $\xi$ are real and imaginary parts of $\nu$:
$\nu=\zeta+i\xi$.
Just form this expression one can see an essential qualitative difference
in dynamics of the AL model  and GDNLS equation. In the first case
\cite{CKVV}, i.e. at $\epsilon=1$,  there is a region of 
parameters (at $|\nu|> 1$) where the lattice becomes unstable, this 
being displayed by an infinite growth of $\xi$ during finite time. In 
contrast, even a small value of the difference $1- \epsilon$ will 
prevent this growth.  The equations for ($\zeta,\xi$) can be obtained 
from Hamilton's equations with respect to the nonstandard Poisson 
brackets \begin{equation} \label{c1} \{ 
f,g\}=[1-(\zeta^2+\xi^2)]\left\{ \frac{\partial f}{\partial\zeta} 
\frac{\partial g}{\partial \xi} - \frac{\partial f}{\partial\xi} 
\frac{\partial g}{\partial\zeta}\right\} \end{equation}
with respect to the Hamiltonian $I_{-1,1}$. Indeed, for $H=I_{-1,1}$
we have
\begin{equation}
\label{c2}
\dot{\zeta}=
\{H,\zeta\}=
-2\frac{1-\epsilon}{\epsilon}
[1-(\zeta^2+\xi^2)]\xi
\end{equation}
\begin{equation}
\label{c3}
\dot{\xi}=
\{H,\xi\}=
2\frac{1-\epsilon}{\epsilon}
[1-(\zeta^2+\xi^2)](\Lambda+\zeta)
\end{equation}
where $\Lambda=\frac{\epsilon}{1-\epsilon}\cos\vartheta$.

Taking into account the explicit form of $I_{-1,1}$ one finds a
convenient parametrization of the problem as
\begin{equation}
\label{c4}
\zeta +\Lambda=R\cos\chi,\,\,\,\,\,\,\,
\xi=R\sin\chi,
\end{equation}
where $R$ is a positive constant playing the role of the radius
of the orbit in the phase portrait and is related to the energy by
\begin{equation}
\label{c5}
R^2=\frac{\epsilon}{1-\epsilon}I_{-1,1}+\Lambda^2-1.
\end{equation}
The dynamical equations (\ref{c2}), (\ref{c3}) are easily solved
and give the following two different types of orbits.

(a) If $R<|1-\Lambda|$ or $R>|1+\Lambda|$ (hereafter without restriction
of generality we take $\phi\in [-\pi/2, \pi/2]$) the motion is
periodic and is governed by the equations
\begin{equation}
\label{e8}
\zeta =R\frac{A\cos\Theta-1}{A-\cos\Theta}+\Lambda,
\end{equation}
\begin{equation}
\label{e9}
\xi= R\sqrt{A^2-1}\frac{\sin\Theta}{A-\cos\Theta},
\end{equation}
where
\begin{equation}
\label{c6}
\Theta =4\sqrt{A^2-1} R\cos\vartheta\cdot t+2\arctan\left(
\frac{\sqrt{A^2-1}}{A+1}
\frac{\xi_0}{R+\Lambda+\zeta_0}\right)
\end{equation}
and
\begin{equation}
\label{c7}
A=\frac{1-R^2-\Lambda^2}{2\Lambda R}\,\,\,\,\,\,\,\,(A^2>1).
\end{equation}
Here $\xi_0, \zeta_0$ denote the  initial values at $t=0$.

(b) If $|1-\Lambda|<R<|1+\Lambda|$ the system displays aperiodic motion
described by the equations
\begin{equation}
\label{e10}
\zeta =-R\frac{A\cosh\Theta-1}{\cosh\Theta+A}+\Lambda
\end{equation}
\begin{equation}
\label{e11}
\xi= R\sqrt{1-A^2}\frac{\sinh\Theta}{\cosh\Theta+A}
\end{equation}
where
\begin{equation}
\Theta= 4R\sqrt{1-A^2}\cos\vartheta \cdot t +2\ln
\frac{
\xi_0\sqrt{1-A^2}+(A+1)(R+\Lambda+\zeta_0)}{
\xi_0\sqrt{1-A^2}-(A+1)(R+\Lambda+\zeta_0)}
\end{equation}
and the constant $A$ is given by (\ref{c7}) (note that now $A^2<1$).

The existence of two regimes oscillating  and aperiodic is evident
from the phase portrait which  consists of circles of the radius R
which in the integrable limit, $\epsilon=1$,  degenerates into a
straight line.
In the aperiodic case a circular trajectory always crosses
the unit circle (given by $|\nu|=1$), the crossing point being
just the singular point. Thus, for $|A|<1$
an initial condition will move on the circle
characterized by R and, as $t\rightarrow\infty$,
it will reach  the singular point on the unit circle.
For $|A|>1$ there will be no crossing and the point will
continue to  rotate on the respective circle.

From the above analysis it follows that if initially
$|q_n|\leq \epsilon^{-1}$ (or $|q_n|\geq \epsilon^{-1}$) this will
be true for all times. This property is valid also for the
integrable AL system.

\section{Small-amplitude dark solitons near singular points}

Rather complete description
of the  dynamics of (\ref{e1}) is also available for the small amplitude pulses.
Indeed
starting with the case of the excitations slightly deviating from
 the singular points, we can consider backgrounds for the nonlinear
excitations in the from of arbitrary (i.e. with arbitrary $k$ values
in the BZ) set of the singular points. For the sake of simplicity in the
present section
we restrict the analysis either to the center of  BZ,
$k=0$, or to the edge of BZ, $k=\pi$.

The AL dark soliton ($\epsilon=1$) is written in the form
\begin{equation}
\label{e2}
q_n^{AL}(t)=\rho
\left[1-\frac{\sin^2\vartheta}
{\cosh^2(Kn-\Omega t)}\right]^{-1/2}
\exp\left[-i\tan\vartheta\tanh(Kn-\Omega t)\right]
\end{equation}
with the parameters $K,\vartheta,\rho$ linked by
\begin{equation}
\label{e3*}
\sinh \frac{K}{2}=\frac{\rho}{\sqrt{1-\rho^2}}\sin\vartheta.
\end{equation}
The phase $\vartheta$ parametrizes the family of the soliton solutions
and is chosen in the interval $[-\pi,\pi]$. The small amplitude limit
corresponds to the case $\vartheta\ll 0$.
In order to expand  (\ref{e2}) around $\vartheta=0$ we observe that the most
natural representation of the solution of the GDNLS equation at $\epsilon \leq
1$ is
\begin{equation}
\label{a1}
q_n=\kappa^n\epsilon^{-1/2}(1-\gamma^2 \mu a_n)e^{-i\gamma\mu\chi_n+i\omega_0
t}
\end{equation}
where $\gamma\ll 1$ is a small parameter,  $a_n=a_n(\tau)$ and
$\chi_n=\chi_n(\tau)$ are two real functions of the slow time
 $\tau=2\gamma\sqrt{2(\epsilon^{-1}-1-\kappa)} t$, and
$\omega_0$ is given by
(\ref{add3}). The parameter
$\kappa=\pm 1$ introduced in (\ref{a1}) will be used to choose two different
background oscillations i.e. in-phase oscillations ($\kappa=1$) and
out-of-phase oscillations ($\kappa=-1$) corresponding
to $k=0$ and $k=\pi$, while the parameter $\mu=\pm 1$ is used to obtain either
dark ($\mu=1$) or bright ($\mu=-1$) pulses.

Substituting (\ref{a1}) in (\ref{e1}) and gathering all the terms  of
orders up to $\gamma^3$  we arrive to the following Toda system \cite{FT}
\begin{equation}
\label{a2}
\frac{d a_n}{d\tau}=a_n(c_n-c_{n-1}),
\end{equation}
\begin{equation}
\label{a3}
\frac{d c_n}{d\tau}=a_{n+1}-a_n,
\end{equation}
where
\begin{equation}
\label{add1}
c_n=\sqrt{2|\epsilon^{-1}-1-\kappa|}(\chi_n-\chi_{n+1}),
\end{equation}
and
\begin{equation}
\label{add2}
\mu=\kappa \mbox{sign} (\epsilon^{-1}-1-\kappa).
\end{equation}
(For the integrable case this system was also  obtained in \cite{CKV}).
We have therefore that  small amplitude dark pulses near the singular
points can be viewed as exact solitons of the Toda lattice moving
on the oscillating background $\rho\exp(i\omega_0t)$. When $\kappa=1$ ($k=0$), it
follows from (\ref{add2}) that $\mu=1$ i.e. the respective solution
(\ref{a1}) is always dark, i.e. its energy profile is always below the
background level. On the other hand, at the edge of the BZ we
have that
$\kappa=-1$ and the sign of
$\mu$ depends on $\epsilon$. We find that $\mu=1$ for $\epsilon<1/2$ and  $\mu=-1$ for
$\epsilon>1/2$. This implies that at $k=\pi$
small amplitude dark solitons exist for $\epsilon>1/2$ while for
$\epsilon>1/2$ bright pulse should appear.
To check these predictions we have  numerically integrated the GDNLS system
with a fifth-order Runge-Kutta adaptive stepsize algorithm (the numerical
error was controlled by checking the conservation of the norm
up to the fifth decimal digit).
In Fig. 1 the time evolution of a small amplitude dark
soliton  on a line of 1500 sites with an in-phase ($\kappa=1$) background
is reported. Here the initial condition is of the
form of a one-soliton solution of
the Toda lattice which corresponds to a bisoliton solution
of the GDNLS.
We see that the initial  pulse splits in two dark
excitations which are stable over
long time. This is in contrast with what happens
for a  bright profile on the same background and for the same
parameter values as reported in Fig.1.
This behavior  is found to be true
for all values of $\epsilon$ in the range $(0,1]$.
The same analysis,  but for an out-of-phase background
($\kappa=-1$), shows that bright pulses are stable and dark ones are unstable
if $\epsilon<1/2$, while dark pulses are stable and bright ones are unstable
if $\epsilon>1/2$. In Figs. 2,3, we have reported the evolution of a
 bright and dark GDNLS bisoliton  for $\epsilon$ values
respectively at $\epsilon=0.25$ and $\epsilon=0.75$. We see that
the pulse are quite stable in agreement with our predictions.
From the above analysis it is also clear that at the "critical" value
$\epsilon =1/2$  the solution changes stability and the dynamics
manifests a strong dispersive character (weakly nonlinear
or effectively linear regime). In  Figs. 4,5  the time evolution of
a  bisoliton, respectively dark and
bright, at  the "critical" value $\epsilon=1/2$ is reported. We see that
in both cases the profiles  decay in background radiation
in agreement with our prediction.

\section{Small-amplitude excitations far from the singular points and shock
waves of the GDLNS}

In order to find dynamical equations for small-amplitude pulses far
from the singular points we apply the same analysis as in the previous
section but now we expand around
\begin{equation}
\label{b4}
q_n=[\rho+a_n(t)]e^{-i\Phi_n(t)},
\end{equation}
where
\begin{equation}
\label{b5}
a_n(t)=
\gamma^2a^{(0)}(T,X;\tau)+
\gamma^4a^{(1)}(T,X;\tau)+...
\end{equation}
\begin{equation}
\label{b6}
\Phi_n(t)=
\gamma\Phi^{(0)}(T,X;\tau)+
\gamma^3\Phi^{(1)}(T,X;\tau)+...
\end{equation}

In the above expansion
$X=\gamma n$, $ T=\gamma t$ denote fast space and time variables while
$\tau =\gamma^3 t$ represents a slow time.
By analyzing the equations of all orders from $\gamma$ up to
$\gamma^5$ we find that at the leading orders ($\gamma^2$, $\gamma^3$)
the GDNLS system
reduces to
\begin{equation}
\label{b7}
\frac{\partial\Phi^{(0)}}{\partial T}=4\rho^2 a^{(0)}
\end{equation}
\begin{equation}
\label{b8}
\frac{\partial a^{(0)}}{\partial T}=
(1-\epsilon\rho^2) \frac{\partial^2 \Phi^{(0)}}{\partial X^2}.
\end{equation}

From  system (\ref{b7}), (\ref{b8}) it readily follows that the
excitation at leading order moves with the velocity
\begin{equation}
\label{k3}
V_{\epsilon}=-2\rho\sqrt{1-\epsilon\rho^2}
\end{equation}
[the given sign of the velocity is chosen only for the sake of
convenience]
which depends on the background level.

 It is of interest to note that
the  velocity $V=\Omega/K$ of the AL dark soliton
 in the small amplitude limit, takes the form
\begin{equation}
\label{k1}
V=V_0+\frac{\vartheta^2 \rho}{3\sqrt{1-\rho^2}}(3-4
\rho^2),
\end{equation}
where  $V_0$ is given by (\ref{k3}) at $\epsilon=1$ and coincides with
 the velocity of  linear waves against the background $\rho$ in the long
wavelength limit (i.e. in the center of BZ). It also
follows from (\ref{k1}) that $V$ coincides with the velocity of
the linear waves  when $\rho^2=3/4$.

In the next two orders in $\gamma$ we obtain the  equations
\begin{equation}
\label{b9}
\frac{\partial\Phi^{(1)}}{\partial T}-4\rho^2 a^{(1)}=-\frac{\partial
\Phi^{(0)}}{\partial \tau}-(1-\epsilon\rho^2)\frac{\partial^2 a^{(0)}}{\partial
X^2}+6\rho^2a^{(0)2},
\end{equation}
\begin{eqnarray}
\label{b9*}
\frac{\partial a_{n}^{(1)}}{\partial T}-
(1-\epsilon\rho^2) \frac{\partial^2 \Phi^{(1)}}{\partial X^2}
\nonumber \\ =-\frac{\partial a^{(0)}}{\partial\tau}+\frac{1}{12}\frac{\partial^3
a^{(0)}}{\partial T\partial
X^2}-(2-5\epsilon\rho^2)\left(\frac{4\rho^2}{V_{\epsilon}}\right)a^{(0)}
\frac{\partial a^{(0)}}{\partial X}.
\end{eqnarray}

It is remarkable that the compatibility condition of these two
equations directly leads to the following KdV equation
\begin{eqnarray}
\label{b10}
-4\rho\sqrt{1-\epsilon\rho^2}\frac{\partial
a^{(0)}}{\partial\tau}-\frac{1}{3}(1-\epsilon\rho^2)[3-(3\epsilon+1)\rho^2]
\frac{\partial^3 a^{(0)}}{\partial Z^3}+8\rho^2(3-4\epsilon\rho^2)
a^{(0)}
\frac{\partial a^{(0)}}{\partial Z}=0,
\end{eqnarray}
where $Z$ denotes the running variable $Z=X-V_{\epsilon}T$.
We remark that a similar result was also obtained in refs.
\cite{KS}.

Equation (\ref{b10}) allows soliton solutions corresponding both to positive
and negative $a^{(0)}$ depending on the sign
of the prefactors in the dispersive
and nonlinear terms. The results of the respective
analysis are illustrated in Fig.6.
In this figure the region between the curves 1 and 2
marked with B corresponds to parameter values for which stable
propagation of bright solitons against the background is possible.
In this region dark solitons will not be stable and they will decay
 in background
radiation.
The line 3 itself determines the amplitude of the singular points:
$\epsilon=1/\rho^2$. Close
to this line the dynamics of small amplitude excitations follows
Toda equations as described in section 2. The
areas marked with D in Fig. 6 correspond to parameter regions
in which  dark solitons are stable while bright ones are unstable.
On the curve 2 of Fig. 6
[$\epsilon=3/(4\rho^2)$]  the nonlinear term becomes zero as it is evident
from (\ref{b10}). This implies that any
localized pulse will spread out in background radiation
(no soliton-like excitations).
On the other hand on curve 1 of Fig. 6, which represents the other
dark-bright interface, the system
has different dynamical properties since it becomes effectively
dispersionless as it is seen from (\ref{b10}).
Since the nonlinearity on this curve is not zero we expect this
case to give shock solutions (see below).

To check these results we have numerically computed
the time evolution of small amplitude bright and dark
excitations for parameter values
taken in different regions of Fig. 6. In Fig. 7a the evolution
of a dark soliton at  $\epsilon=0.4, \rho=0.8$ is reported.
In Fig.7b we show the same evolution but for an initially
bright pulse for the same parameter values as in Fig.7a.
From these figures it is clear that parameter values on the left
of  curve 1 correspond to stable dark pulse
propagation and to decaying
 bright excitations. A similar behavior is found in the region
between curve 2 and curve 3. This
is seen from Figs, 8a,b in which the evolution of, respectively, a
dark and a bright excitation is reported at parameter values
$\epsilon=0.4, \rho=1.5$.
In the region between curve 1 and curve 2 the behavior is just the opposite
i.e. dark solitons decay in background radiation
while bright pulses can propagate as solitary waves.
This is shown in Figs. 9 a,b for
respectively a bright and a dark initial profile at parameter values
$\epsilon=0.16, \rho=1.7$. It is of interest to remark that while
for the pulses
of Fig. 7b,  the radiation is on the front in Figs. 8b,9b
the radiation is just on the back. This corresponds to the fact that
wave packets in the region on the left of the curve 1
always have group velocity greater than the velocity $V_{\epsilon}$ of the
background radiation while the opposite is true for the complementary regions.

From (\ref{b10}) it is evident that for $\epsilon=1/\rho^2-1/3$
the dispersion disappears and the resulting equation
possesses shock wave solutions. This surprising property of the GDNLS system
has been numerically checked in Fig. 10. In this figure the time
evolution of an initial bright profile is reported for parameter values
taken on the curve 1 of Fig.6 at  $\epsilon=0.361(1)$, $\rho=1.2$.
We see that the smooth initial profile is distorted with the top moving
at higher velocity of the bottom part of the profile. This give rise to a
forward bending of the profile with breaking of the wave and rapid
oscillations on the wave front.
 A similar behavior is observed also
for dark pulses as it will be reported in  more extended form elsewhere
\cite{ref}.
We should mention that the possibility of
shock waves in other chains was also reported  in ref.
\cite{toda}.

\section{Conclusion}

We have analyzed the properties of the small-amplitude solutions of the GDNLS.
 Like the well-known AL model the GDNLS equation  has  singular points for all
$\epsilon$ different from zero which
provide effective decoupling of the chain.
In this sense one can speak about dominant role of the nonlinearity of the
"AL-type" in the behavior of the system. Meantime, the properties of the
singular points are essentially different in the cases $\epsilon=1$ and
$\epsilon<1$.

The small amplitude excitations against   constant amplitude backgrounds
are governed by the Toda-lattice equation if they are in the vicinity of the
singular points or by the KdV equation if the background amplitude is less
than $1/\sqrt{\epsilon}$. The type of the solution essentially depends not
only on the deformation parameter but also on the chosen point in the BZ.
Existence of the bright pulses against nonzero background  above
singular points  is a nontrivial qualitative demonstration of the mentioned
differences. We have delimited the  regions in the parameter space in which
dark solitons are stable in contrast with regions in which
bright pulses on  nonzero background are. On the
boundaries of these regions we have found, quite surprisingly,
 that  shock waves  may exist.

\section{Acknowledgment}

VVK thank the Deparment of Theoretical Physics of the  University of Salerno,
 for the warm hospitality. Financial
support from INTAS grant No. 93-1324 and from INFM unita' di Salerno is also
acknowledged.

\figure{ Time evolution of an initial dark pulse ($\mu=1$)
 on an in-phase background
($\kappa=1$) at $\epsilon=0.25$. The total integration time is 300.
The initial condition corresponds to a soliton of the Toda lattice.}

\figure{ Time evolution of an initial bright ($\mu=-1$) pulse  on
an out-of-phase background
($\kappa=-1$) at $\epsilon=0.25$. The total integration time is 300.
The initial condition corresponds to a soliton of the Toda lattice.}

\figure{ Same as in Fig.2 but for a dark initial condition at
$\epsilon=0.75$.}

\figure{ Time evolution of an initial dark pulse   ($\mu=1$) on
an out-of-phase background ($\kappa=-1$) at the critical value
$\epsilon=0.5$. The total integration time is 1200.}

\figure{ Same as in Fig.4 but for a bright pulse  ($\mu=-1$).}

\figure{ Parameter space ($\epsilon,\rho$).  The window
where bright pulses against a
nonzero background exist is bounded by the curves (1)
$\epsilon=1/\rho^2-1/3$ and (2) $\epsilon=3/(4\rho^2)$.
The curve (3), $\epsilon=1/\rho^2$, corresponds to the singular points and in its vicinity the
small-amplitude excitations of the system are governed by the Toda-lattice
equation.}

\figure{ (a) Time evolution of an initial dark  pulse
 corresponding to a KdV
 soliton  for parameter values $\epsilon=0.4, \rho=0.8$. (b) Same as in (a) but
for  initial bright pulse. The total integration time is $T=1200$.}

\figure{(a) Time evolution of an initial dark pulse  for parameter values
$\epsilon=0.4, \rho=1.5$.
(b) Same as in (a) but for an initial bright pulse.
 The total integration time is $T=1200$.}

\figure{(a) Time evolution of an initial bright pulse  for parameter values
$\epsilon=0.16, \rho=1.7$.
(b) Same as in (a) but for an initial dark pulse.
 The total integration time is $T=1200$.}

\figure{ Shock wave formation from a smooth bright pulse at parameter values
$\epsilon=0.361(1), \rho=1.2$ on curve 1 of Fig. 6.
 The total integration time is $T=1200$.}

\end{document}